\documentstyle[12pt]{article}
\input epsf

\newcommand{\sect}[1]{\setcounter{equation}{0}\section{#1}}

\newcommand\M{{\cal M}}

\newcommand\R{{\mathrm {I\!R}}}

\newcommand{\ap}{Ann. Phys.}
\newcommand{\cqg}{Class. Quantum Grav.}
\newcommand{\cmp}{Comm. Math. Phys.}
\newcommand{\grg}{Gen. Rel. Grav.}
\newcommand{\jmp}{J. Math. Phys.}

\newcommand{\pr}{Phys. Rev.}
\newcommand{\prl}{Phys. Rev. Lett.}

\newcommand{\ra}{\rightarrow}
\newcommand{\lra}{\leftrightarrow}

\newtheorem{lemma}{Lemma}
\newtheorem{theorem}{Theorem}

\newcommand{\insertfig}[2]{\leavevmode \vcenter{\hbox{\epsfxsize=#2 cm
\epsffile{#1.ps}}}}

\begin{document}

\begin{titlepage}
%\vspace{.5in}

\begin{flushright}
DAMTP-97-127\\
November 1997\\
gr-qc/9711069
\end{flushright}

\vspace{.5in}

\begin{center}
{\LARGE \bf
Building blocks for topology change in 3D}\\

\vspace{.4in}
{\large {Radu Ionicioiu}
        \footnote{\it on leave from Institute of Gravitation and
        Space Sciences, 21-25 Mendeleev Street, 70168 Bucharest, Romania}\\
       {\small\it DAMTP, University of Cambridge}\\
       {\small\it Silver Street, Cambridge, CB3 9EW, UK}\\
       {\small\tt email: ri10001@damtp.cam.ac.uk}\\}
\end{center} 

\vspace{.5in}
\begin{center}
{\large\bf Abstract}
\end{center}

\begin{center}
\begin{minipage}{4.75in}
{\small
We investigate topology change in 3D. Using Morse theory and handle decomposition we find the set of elementary cobordisms for 3-manifolds. These are: (i) $\O \lra S^2$; (ii) $\Sigma_g \lra \Sigma_{g+1}$; (iii) $\Sigma_{g_1} \sqcup \Sigma_{g_2} \lra \Sigma_{g_1+g_2}$, and they have appealing physical interpretations, e.g.~Big Bang/Big Crunch, wormhole creation/annihilation and Einstein-Rosen bridge creation/annihilation, respectively. This decomposition into building blocks can be used in the path integral approach to quantum gravity in the sum over topologies.

~~
}
\end{minipage}
\end{center}

\end{titlepage}
\addtocounter{footnote}{-1}

\sect{Introduction}

Topology change has become recently a subject of increasing research interest \cite{gibb1, gibb2, sorkin2, madore}. The image of fluctuating topology at the Planck scale is due to Wheeler, who was the first one to point out the dynamical topology inherent to that scale, the now famous {\em foamlike structure} of spacetime \cite{wheeler}.

Although from a classical point of view topology change is excluded \cite{geroch}, in the quantum case this is different due to fluctuations of the metric. There are many arguments in favour of topology change. In a sum over histories approach to quantum gravity, the sum over metrics is naturally extended to a sum over topologies. Another argument comes from the Big Bang, which implies a topological transition $\O \ra S^3$. Our conclusion is that topology change becomes an essential ingredient of Planck scale physics.

In this article we investigate topology change by using Morse theory and handle decomposition. Applied to the case of 3-manifolds, this yields the set of {\em building blocks}, i.e. those elementary cobordisms from which any 3-fold can be built (up to a homeomorphism).

While finalising this article, reference \cite{garcia} came to our attention, in which a similar framework of handle decomposition was used for topology change.

\vspace{.5cm}

From the beginning, it is important to make a distinction between a topological and a Lorentzian cobordism \cite{yodzis, reinhart}.

By a {\em topological cobordism} we understand a smooth, compact, $n$-dimensional manifold $\M$ whose boundary has 2 disjoint components $\partial \M= \M_0 \sqcup \M_1$, with $\M_0$ and $\M_1$ two smooth, closed, $(n-1)$-dimensional manifolds (possible empty or nonconnected). Two manifolds are topologically cobordant if and only if they have the same Stiefel-Whitney and Pontrjagin numbers (the oriented case) or only the Stiefel-Whitney numbers (the non-oriented case) \cite{wall, milnor}.

A {\em Lorentzian cobordism} is a topological cobordism $(\M;\ \M_0, \M_1)$ together with a nonsingular vector field ${\bf v}$ which is interior normal to $\M_0$ and exterior normal to $\M_1$. In this case we can define a nonsingular Lorentz metric $g^L_{\mu \nu}$ on $\M$

\begin{equation}
g^L_{\mu \nu} = g^R_{\mu \nu} - \frac{2 v_\mu v_\nu}{g^R_{\alpha \beta} v_\alpha v_\beta}
\label{glr}
\end{equation}
where $g^R_{\mu \nu}$ is a Riemann metric on $\M$. This is always possible, since there is a one-to-one correspondence between nonsingular vector fields ${\bf v}$ and Lorentz metrics $g^L_{\mu \nu}$ on $\M$. With respect to $g^L_{\mu \nu}$, $\M_0$ and $\M_1$ are spacelike and we will denote them as the initial and final hypersurfaces of the cobordism.

\vspace{.5cm}

Following a celebrated theorem of Geroch \cite{geroch}, topology change implies either {\em closed timelike curves} (CTCs) or {\em singularities} in the metric. For the rest of this article we assume there are no CTCs, and therefore we admit singularities in the metric, in order to have topology change. A consequence of the other choice (CTCs, no singularities) for topology change in Kaluza-Klein theories has been studied in \cite{rikk}.

The first question we ask is: How serious are such singularities? First of all, we have to point out that these are not {\em curvature singularities} (like $r=0$ in the Schwarzschild metric). In our case spacetime is a smooth manifold and therefore the curvature is bounded. However, at the singular points the metric $g_{\mu \nu}$ fails to be invertible, this being related to the singularities in the vector field $\bf v$ which defines the 'time flow' in (\ref{glr}). As Horowitz pointed out \cite{horowitz}, if we allow degenerate tetrads, the singularities can be very mild, since the curvature is bounded.

The viewpoint adopted here is the following: until we have a full theory of quantum gravity we shall leave all the options open, and therefore singular metrics are a legitimate object to study.

The important result is \cite{sorkin}:\\
~~\\
{\em Every topological cobordism admits a metric which is Lorentzian everywhere, except for a finite number of singularities}.\\
~~\\
In order to study the singularities associated with the topology change, we need a few standard results from Morse theory.

\sect{A brief introduction to Morse theory}

In what follows we review some standard results in Morse theory. More details can be found in \cite{miln1, miln2, miln3}.

The essence of Morse theory is to study the topology of a manifold $\M$ by analysing the critical points of a smooth, real function \mbox{$f:\M \ra \R$}.\\
~~\\
{\bf Definition:} Let $f:\M \ra \R$ be a smooth function and $dim \M=n$. A point $p \in \M$ is a {\em critical point} of $f$ if, in some coordinate system we have

\[ \left. \frac{\partial f}{\partial x_1} \right|_p = \ldots = \left. \frac{\partial f}{\partial x_n} \right|_p = 0 \]
and $p$ is a $\em nondegenerate$ critical point if

\[ det\left( \left. \frac{\partial^2 f}{\partial x_i \ \partial x_j}\right|_p \right) \neq 0 \]
A {\em critical value} $c_i \in \R$ is the image of a critical point, $c_i= f(p_i)$. For the rest of this article all critical points will be assumed to be nondegenerate.

The key result is the Morse Lemma.

\begin{lemma}[Morse] If $p \in \M$ is a nondegenerate critical point of $f$, then

\[ f(x_1 \ldots x_n)= f(p) -x^2_1- \ldots -x^2_\lambda +x^2_{\lambda+1}+ \ldots + x^2_n\]
in some coordinate system in a neighbourhood $U(p)$.
$\lambda$ is called the {\em index} of the (nondegenerate) critical point $p\in \M$.
\end{lemma}

Let $(\M; \M_0, \M_1)$ be a cobordism. A smooth, real function \mbox{$f:\M \ra [a,b]$} is a {\em Morse function} if:\\
1) $f^{-1}(a)= \M_0$\\
2) $f^{-1}(b)= \M_1$\\
3) all the critical points $p_i$ of $f$ are interior (i.e. $p_i \notin \partial \M$) and nondegenerate.

Without loss of generality, we can assume that the critical values of $f$ are distinct, $p_i \neq p_j \Rightarrow c_i=f(p_i)\neq f(p_j)=c_j$ (i.e., the Morse function is {\em proper}).

The {\em Morse number} of a cobordism $\mu(\M)$ is the minimum (over all the Morse functions defined on $\M$) of the number of critical points,

\[ \mu(\M) = min_f\{\# critical\ points\ of f\ | \ f- Morse\} \]

Thus, the sphere has $\mu(S^3)=2$, the cylinder $\mu(\Sigma \times I)=0$ and the torus $\mu(T^2)=4$.

We have the following general result \cite{miln1, miln3}:

\begin{theorem}

(i) Every cobordism has a Morse function.

(ii) A Morse function has a finite number of critical points.

(iii) If $f:\M \ra \R$ is a Morse function with no critical points, then $\M$ is topologically trivial, $\M \cong \Sigma \times [0, 1]$.
\end{theorem}

\noindent
{\em Observation:} The converse of (iii) is, however, not true. A topologically trivial manifold (e.g. $\Sigma \times I$) can have a nontrivial Morse function (i.e. with critical points). An example is an U-shaped cylinder which has 2 critical points (the Morse function is the height function from a plane tangent to bottom of the cylinder).

As a consequence of (iii), topology change can occur only at the critical points. Since the number of critical points is finite and the critical values are distinct, we can 'slice up' the manifold between the critical points. Thus, any cobordism can be expressed as a composition of cobordisms with Morse number $\mu=1$, these being the {\em building blocks}.
The idea of introducing building blocks for topology change has been also studied by Alty \cite{alty} in the context of 4-dimensional manifolds, but without giving an explicit construction of them.

Next, we focus on the structure of the elementary cobordisms (i.e. those with Morse number $\mu=1$).

\sect{Attaching handles}

A cobordism $(\M; \M_0, \M_1)$ which has a Morse function with a single critical point of index $\lambda$ is called an {\em elementary cobordism of index $\lambda$} (or shortly, a {\em $\lambda$-cobordism}). Obviously, a $\lambda$-cobordism is an elementary building block.

In any dimension $n$, there can be only $n+1$ types of (nondegenerate, or Morse) critical points, since $\lambda=0 \ldots n$. Moreover, a $\lambda$-cobordism is homeomorphic to one of the $(n-\lambda)$-cobordisms (since for given $\lambda$, there can be several $\lambda$-cobordism which are not homeomorphic-- see below the case of 1-cobordisms in 3 dimensions). This can be easily seen from the following argument. If $f$ has a critical point $p$ of index $\lambda$, then for the Morse function $g=-f$, $p$ is a critical point of index $n-\lambda$. Thus the $(n-\lambda)$-cobordism represents the same cobordism as a $\lambda$-cobordism, but 'upside-down', and therefore it mediates the inverse topological transition $\Sigma_{final}\ra \Sigma_{initial}$.

The following is a standard theorem in cobordism theory \cite{miln3,fuks}.

\begin{theorem}
Any cobordism $(\M; \M_0, \M_1)$ can be obtained from the trivial cobordism $\M_0 \times I$ by attaching a finite number of $\lambda$-handles.
\end{theorem}

By {\em attaching a $\lambda$-handle} to a boundary $\M_0$ we understand gluing an $n$-ball $D^n= D^{\lambda} \times D^{n-\lambda}$ via an arbitrary embedding $h: S^{\lambda -1} \times D^{n-\lambda} \ra \M_0$.

In order to find the $\lambda$-cobordisms we start with the cylinder $\M_0 \times I$ over an arbitrary boundary and find all the embeddings $h: S^{\lambda -1} \times D^{n-\lambda} \ra \M_0$ for the boundary of a given $\lambda$-handle. The manifold obtained from the cylinder $\M_0 \times I$ after gluing the $\lambda$-handle along this embedding will be a $\lambda$-cobordism.

\subsection{Handles in 2D}

We start with a 'warm-up' example in 2D. The possible boundary for a 2-manifold is $S^1$ (the only closed 1-dimensional manifold), or a disjoint sum of circles $S^1 \sqcup \ldots \sqcup S^1$.
Attaching a handle in 2D is equivalent to gluing a disk (i.e. a 2-ball) $D^2$ along different parts of its boundary.\\
~~\\
{\bf 0-cobordism}

In this case we have to attach $D^2 = D^0 \times D^2$ via the empty set $S^{-1} \times D^2 = \O$ (since $S^{-1}= \O$).

Starting with a manifold $\M$ with boundary $\partial \M$, attaching a $0$-handle is equivalent to creating an $S^1$ boundary out of nothing and therefore the new boundary will be $\partial \M \sqcup S^1$. Thus, the $0$-cobordism is simply a disk $D^2$ and the topology transition mediated is $\O \ra S^1$.\\
~~\\
{\bf 1-cobordism}

The same disk $D^2 = D^1 \times D^1$ is glued now along $S^0 \times D^1$ (two disjoint line segments).
There are two different embeddings of the two segments $S^0 \times D^1$ in the boundary of a 2-manifold. The two segments can belong either to the same $S^1$ component, or to different $S^1$ components of the boundary. In the first case we have the following figure

$$\insertfig{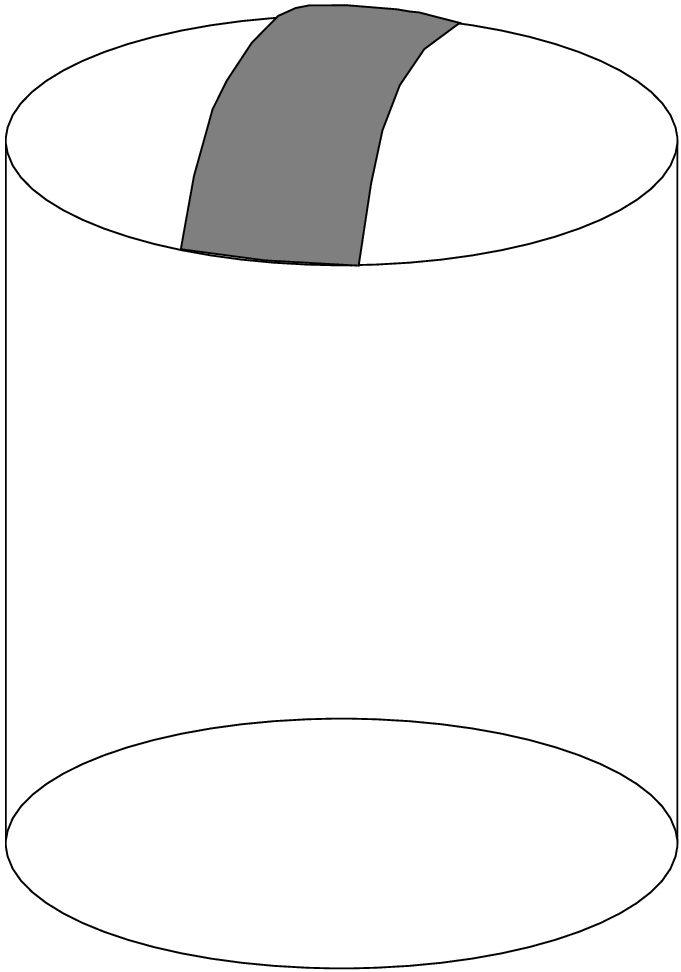}{1} \ \ \cong \ \insertfig{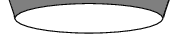}{2}$$

The second case corresponds to:

$$\insertfig{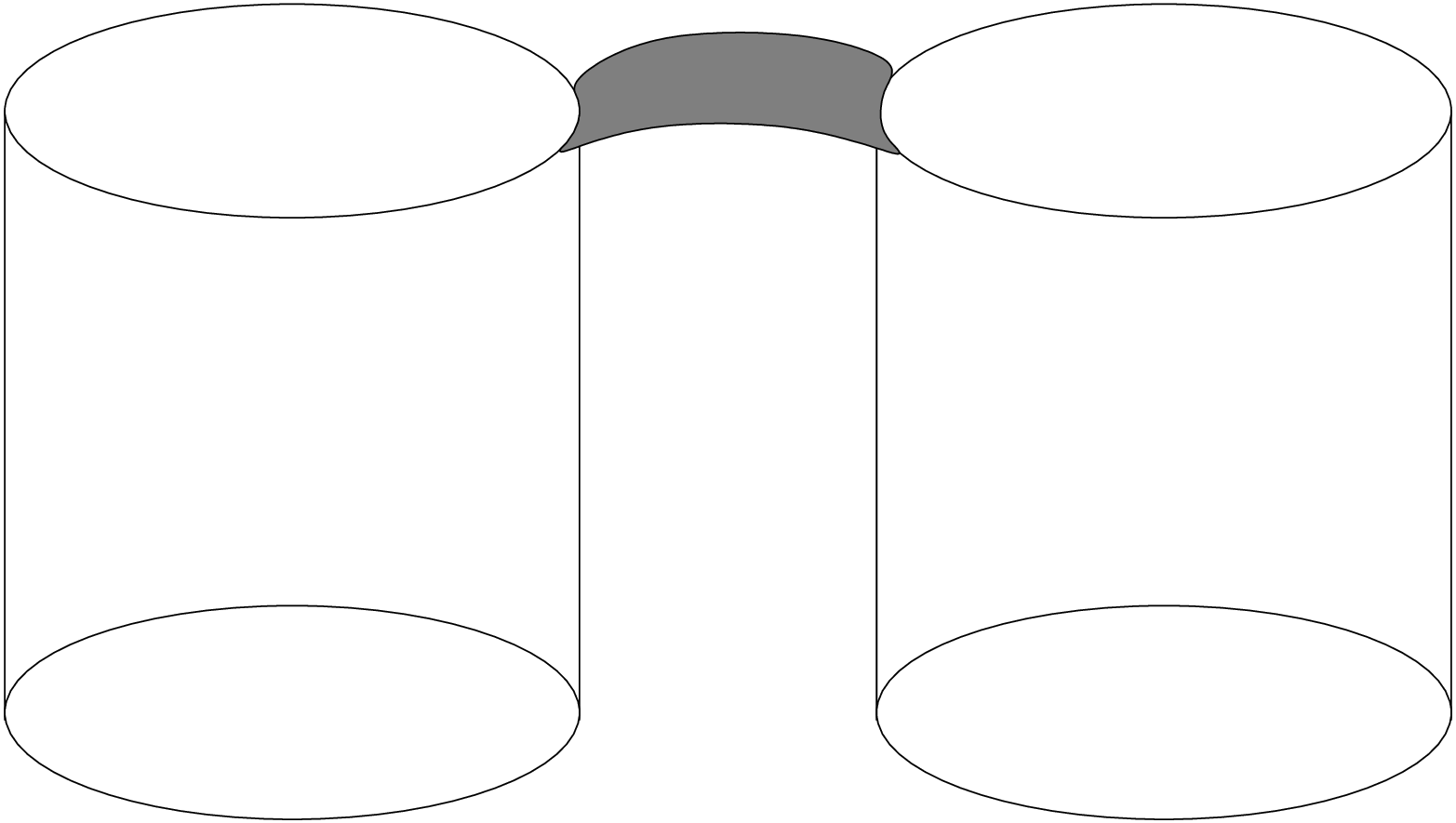}{2.5} \ \ \cong \ \insertfig{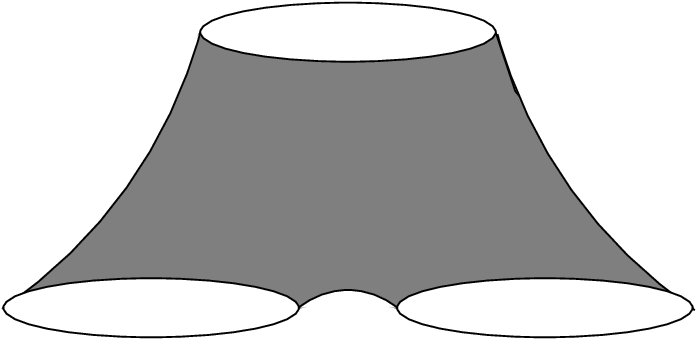}{2}$$

Another possibility is to twist the 'ribbon' $D^2$ before gluing its two ends on the same $S^1$ component; the resulting manifold is the connected sum of a M\"obius band and a disk. Since we are not interested in nonorientable boundaries, we do not consider this case here.\\
~\\
{\bf 2-cobordism}

This is the reverse of a 0-cobordism. The disk $D^2=D^2 \times D^0$ is glued along its whole $S^1 \times D^0= S^1$ boundary ($D^0$ is just a point). Attaching a 2-handle reduces then to gluing a disk to one of the existing $S^1$ boundaries.

In conclusion, the only two elementary cobordisms in 2D are the 'trousers' and the 'Big Bang' (or {\em yarmulke}, see \cite{louko}). Schematically, we have:

$\begin{array}{lccc}
\lambda\\
0\ \ \ +\ +\ \ \ \insertfig{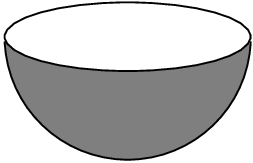}{1}:\ \ \ \ \O \ra S^1\\ \\
1\ \ \ -\ +\ \ \ \insertfig{2d1}{1}:\ \ \ \ \cases{S^1 \ra S^1 \sqcup S^1\cr S^1 \sqcup S^1 \ra S^1\cr}\\ \\
2\ \ \ -\ -\ \ \ \insertfig{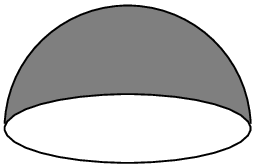}{1}:\ \ \ \ S^1 \ra \O\\
\end{array}$

%%%%%%

\subsection{Handles in 3D}

Now we can do the same analysis for the 3D case. The general boundary of a 3-manifold is homeomorphic to a genus $g$ surface $\Sigma_g$ (or a disjoint sum of such surfaces). We consider only manifolds with orientable boundaries, therefore we exclude from the possible boundaries closed 2-manifolds which have the projective plane ${\bf RP}^2$ as a factor.

In order to construct the $\lambda$-cobordisms we start with the cylinder $\Sigma_g \times I$ and attach to one of the $\Sigma_g$ boundaries a $\lambda$-handle. The cylinder $\Sigma_g \times I$ can be viewed as a hollow genus $g$ handlebody. As a 3-manifold it has two $\Sigma_g$ boundaries -- the exterior one and the interior one, which is shaded in the figure:

$$\insertfig{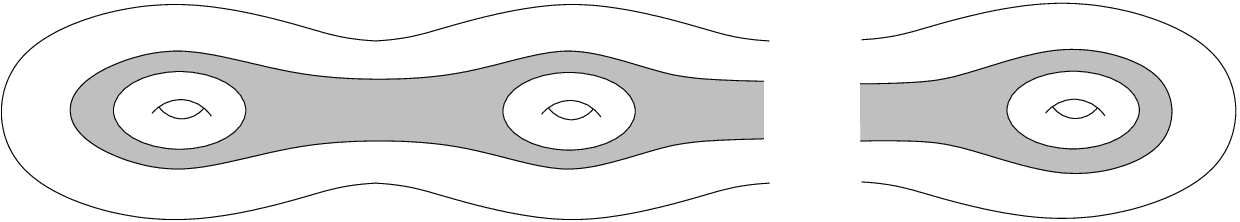}{10}$$\\
~~\\
{\bf 0-cobordism}

This is similar to the 2D case and represents the creation of an $S^2$ boundary out of the vacuum (we glue $D^0 \times D^3$ via the empty set $S^{-1} \times D^3$). The cobordism is just a three-ball $D^3$ which mediates the transition $\O \ra S^2$.\\
~~\\
{\bf 1-cobordism}

Attach the three-ball $D^1 \times D^2$ along $S^0 \times D^2$ (two disks) on an arbitrary $\Sigma_g$ boundary. It is equivalent to gluing a solid tube along its two opposite ends. There are two possible embeddings.

(i) both ends on the same boundary:

$$\Sigma_g \ra \Sigma_{g+1}:$$
$$\insertfig{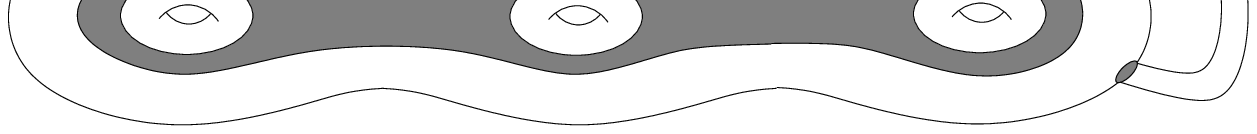}{6} \ \ \cong \ \ \insertfig{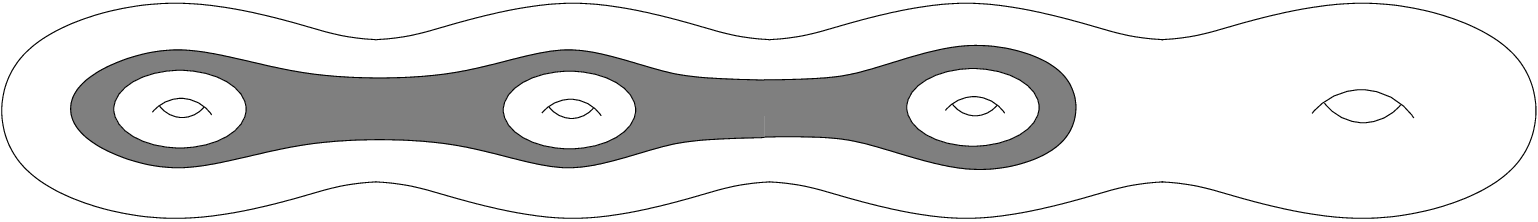}{6}$$ 
~~\\
This is a wormhole creation.

(ii) the two disks glued on disjoint boundaries:

$$\Sigma_{g_1} \sqcup \Sigma_{g_2} \ra \Sigma_{g_1+g_2}:$$
$$\insertfig{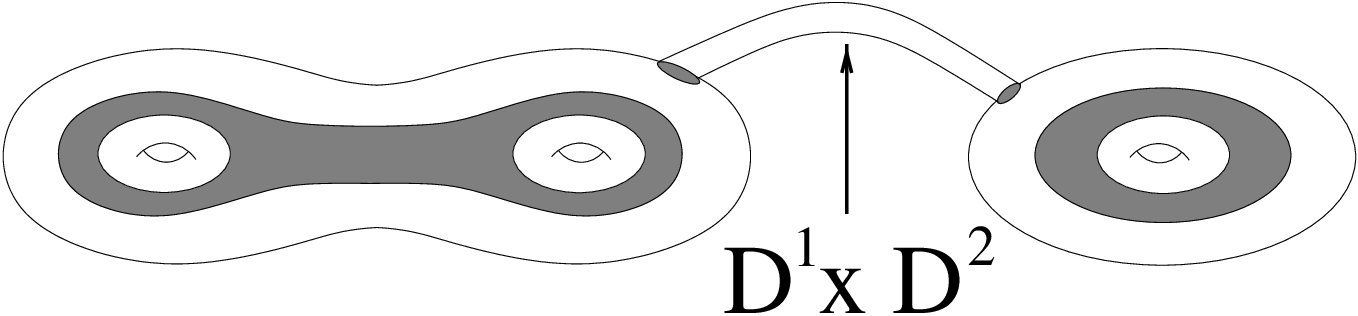}{5} \ \ \ \cong \ \ \ \insertfig{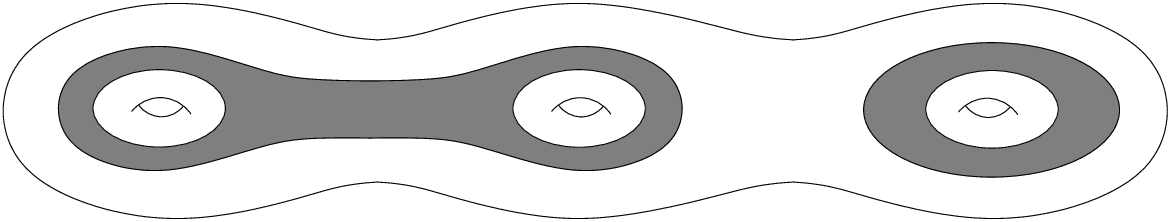}{5}$$
~~\\
This represent the creation of an Einstein-Rosen bridge (connecting two disjoint 'universes').\\
~~\\
{\bf 2-cobordism}

\begin{figure}
\epsffile{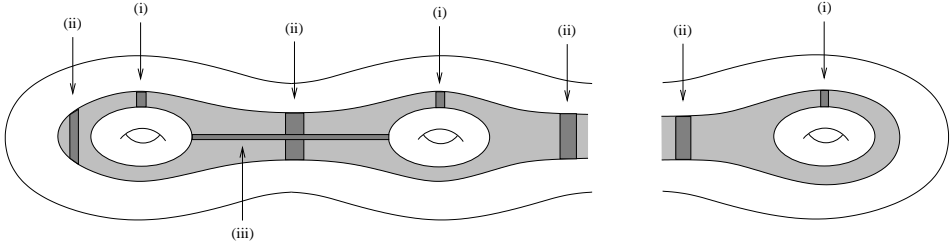}
\caption{Different ways of attaching a 2-handle and the resulting \mbox{2-cobordisms:} (i) $\Sigma_g \ra \Sigma_{g-1}$; (ii) $\Sigma_g \ra \Sigma_k \sqcup \Sigma_{g-k}$, ($k=0 \ldots g$); (iii) this reduces, after a homeomorphism, to case (i), $\Sigma_g \ra \Sigma_{g-1}$.}
\label{3d2}
\end{figure}

This should be equivalent to one of the 1-cobordisms, as we can check. Attach $D^2 \times D^1$ (viewed as a solid cylinder) along its lateral surface $S^1 \times D^1$. We have to find different embeddings of this surface in the $\Sigma_g$ boundary. This can be seen in Fig.~\ref{3d2}, where the gluings are done on the interior $\Sigma_g$ of the cylinder $\Sigma_g \times I$.
The gluings of type (i) sever one of the handles, and thus they are equivalent to $\Sigma_g \ra \Sigma_{g-1}$. Type (ii) gluings separate the inner boundary into two disjoint boundaries, $\Sigma_g \ra \Sigma_k \sqcup \Sigma_{g-k}$, with $k=0 \ldots g$.
Type (iii) reduces to type (i) after a homeomorphism of the $\Sigma_g$ boundary.\\
~~\\
{\bf 3-cobordism}

Simply cap an $S^2$ boundary, $S^2 \ra \O$; the cobordism is again a 3-ball $D^3$.

~~\\
To summarize, we have:\\
$\begin{array}{lccc}
\lambda\\
0\ \ \ +\ +\ +\ \ \  \ \ \ \ \O \ra S^2\\ \\
1\ \ \ -\ +\ +\ \ \  \ \ \ \ \cases{\Sigma_g \ra \Sigma_{g+1}\cr \Sigma_{g_1} \sqcup \Sigma_{g_2} \ra \Sigma_{g_1+g_2} \cr}\\ \\
2\ \ \ -\ -\ +\ \ \  \ \ \ \ \cases{\Sigma_g \ra \Sigma_{g-1} \cr \Sigma_{g_1+g_2} \ra \Sigma_{g_1} \sqcup \Sigma_{g_2} \cr}\\ \\
3\ \ \ -\ -\ -\ \ \  \ \ \ \ S^2 \ra \O\\
\end{array}$

The physical interpretation of these elementary building blocks in 3D is:\\
i) Big Bang/Big Crunch: $\O \lra S^2$\\
ii) wormhole creation/annihilation: $\Sigma_g \lra \Sigma_{g+1}$\\
iii) Einstein Rosen bridge creation/annihilation: $\Sigma_{g_1} \sqcup \Sigma_{g_2} \lra \Sigma_{g_1+g_2}$

A similar approach, but using spherical modifications instead of handle decomposition was used in \cite{yodzis, yodzis2, km}; however, the authors omited the 2-cobordism representing the Einstein-Rosen bridge, $\Sigma_{g_1} \sqcup \Sigma_{g_2} \ra \Sigma_{g_1+g_2}$.\\
~~\\
{\em Observation:} At first sight, it seems that the cobordism $\Sigma_{g_1} \sqcup \Sigma_{g_2} \ra \Sigma_{g_1+g_2}$ is a composite one. Thus, we could try to obtain it from $S^2 \sqcup S^2 \ra S^2$ by applying on each $S^2$ boundary an appropiate number of 'wormhole creation' cobordisms $\Sigma_g \ra \Sigma_{g+1}$. However, this is not so, and a counterexample is the following.

\begin{figure}
\epsffile{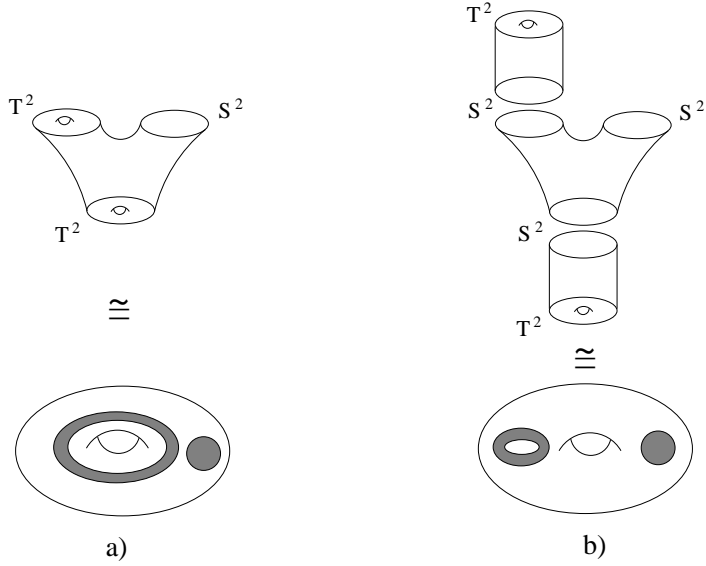}
\caption{Nonhomeomorphic cobordisms (the shaded regions are holes in the solid tori): a) $(T^2 \times I) \# D^3$; b) $(D^2 \times S^1) \# (D^2 \times S^1) \# D^3$}
\label{tts}
\end{figure}

Consider the building block $T^2 \sqcup S^2 \ra T^2$. It is easy to see that this is just the connected sum of the cylinder $T^2 \times I$ and the 3-ball $D^3$ (any $S^2$ boundary can be obtained by taking the connected sum with $D^3$), namely $(T^2 \times I) \# D^3$. On the other hand, we can start with the simplest 'trousers' $S^2 \sqcup S^2 \ra S^2$ and glue on two of the boundaries the 1-cobordism $S^2 \ra T^2$. The resulting manifold is the connected sum of two solid tori and a three-ball, i.e. $(D^2 \times S^1) \# (D^2 \times S^1)\# D^3$. The two cobordisms are not homeomorphic, since $T^2 \times I \not \cong (D^2 \times S^1) \# (D^2 \times S^1)$. This can be checked by computing the Euler characteristics of the two cobordisms, $\chi(T^2 \times I)=0$, whereas $\chi[(D^2 \times S^1) \# (D^2 \times S^1)]= 2\chi(D^2 \times S^1)-2 = -2$. The difference between these two cobordisms is depicted schematically in Fig. \ref{tts} (both are constructed from a solid torus by removing from its interior a ball $D^3$ and another solid torus $D^2 \times S^1$, the difference being in the way the $D^2 \times S^1$ is removed).

\subsection{Handles in $n$ dimensions}

In general, finding the elementary cobordisms in $n$ dimensions is difficult, since it requires at least a classification of closed $(n-1)$-dimensional manifolds, which are the boundaries.
The only manageable cases are the 0 and 1-cobordisms (and their duals, the $n$- and $(n-1)$-cobordisms).
Index 0-cobordism is just a creation of an $S^{n-1}$ boundary, whereas the $n$-cobordism is the 'capping' of an $S^{n-1}$ boundary. Both are equivalent to an $n$-ball $D^n$.

The 1-handle is more complicated. Intuitively, it is similar to the 1-handle of the 3D case. We have to attach the $D^n=D^1 \times D^{n-1}$ along an arbitrary embedding of $S^0 \times D^{n-1}$ (two $(n-1)$-balls). This 1-handle is the higher dimensional analog to the solid tube from the 3D case. Again, the two ends can be on the same $(n-1)$ boundary, say $V^{n-1}$ (wormhole creation), or they can be on two different boundaries (Einstein-Rosen bridge creation). In the first case, this is equivalent to taking the connected sum of the initial boundary with the $n$-dimensional wormhole $S^{n-2} \times S^1$.

The other handles are much more difficult to visualize and will not be studied here.

$\begin{array}{lll}
\lambda\\
0\ \ \ \ \ \ \ \ \ \ \O \ra S^{n-1} \\ \\
1\ \ \ \ \ \ \ \ \ \cases{V^{n-1} \ra V^{n-1}\ \#\ (S^{n-2} \times S^1) \cr V^{n-1} \sqcup W^{n-1} \ra V^{n-1}\ \#\ W^{n-1} \cr}\\
\vdots\\
n-1\ \ \ \thinspace \cases{V^{n-1}\ \#\ (S^{n-2} \times S^1) \ra V^{n-1} \cr V^{n-1}\ \#\ W^{n-1} \ra V^{n-1} \sqcup W^{n-1} \cr}\\ \\
n\ \ \ \ \ \ \ \ \ \ S^{n-1} \ra \O \ \ \ \\
\end{array}$

with $V^{n-1},\ W^{n-1}$ arbitrary closed $(n-1)$-manifolds.
 
\sect{Conclusions}

In this article we analysed the structure of building blocks for topology change in 3 dimensions. The mathematical tools used were Morse theory and handle decomposition.

The advantage of handle decomposition over other forms of constructing 3-manifolds -- e.g. Heegard splitting -- is that it can be applied to any dimension. Moreover, it can be applied also to manifolds with boundary, in contrast to Heegard splitting, which yields only closed 3-manifolds.
Other methods for manifold construction include:\\
i) gluing faces of polyhedra: this can be applied in any dimension, but it lacks a direct intuitive feeling/interpretation;\\
ii) Dehn surgery on knots and links in $S^3$: they can be applied only to (closed) 3-manifolds and it also lacks a simple intuitive picture;

In our opinion, one of the main advantages of handle decomposition over other methods consists in the fact that it can be seen as a 'time evolution' from an initial to a final hypersurface. Each step includes only one elementary topology change (i.e. only one critical point), since each building block has Morse number 1.

~\\
{\Large\bf Acknowledgements}\\
~~\\
This work has been kindly supported by Cambridge Overseas Trust, the Ra\c tiu Foundation and ORS. The author wishes to thank Dr~Ruth Williams and Dr~Dennis Barden for helpful discussions.

\end{document}